\documentclass[10pt,aps,prl,twocolumn,superscriptaddress,notitlepage,nofootinbib]{revtex4-1}
\usepackage{amsmath}
\usepackage{bm}
\usepackage{graphicx}
\usepackage{color,soul}
\usepackage{multirow}
\usepackage{subcaption}
\usepackage[inline]{enumitem}
\def\bea#1\eea{\begin{align}#1\end{align}}
\newcommand{\nn}{\nonumber\\}
\newcommand{\bef}{\begin{figure}[h!tb]\centering}
\newcommand{\eef}{\end{figure}}

\newcommand{\nbar}{{\bar n}} 
\allowdisplaybreaks

\begin{document}
\title{Power Counting the Small-$x$ Observables}

\author{Zhong-Bo Kang }
\email{zkang@physics.ucla.edu}
\affiliation{Department of Physics and Astronomy, University of California, Los Angeles, California 90095, USA}
\affiliation{Mani L. Bhaumik Institute for Theoretical Physics, University of California, Los Angeles, California 90095, USA}

\author{Xiaohui Liu }
\email{xiliu@bnu.edu.cn}
\affiliation{Center of Advanced Quantum Studies, Department of Physics, Beijing Normal University, Beijing 100875, China}

\date{\today}

\begin{abstract}
We emphasize the importance of applying power counting to the small-$x$ observables, which introduces novel soft contributions usually missing and allows for a unified treatment of the Balitsky-Kovchegov (BK) evolution and various Sudakov logarithms. We use $pA \to h(p_{h\perp})X$ at forward rapidity to highlight how the power counting yields a partonic cross section with collinear and soft sectors. We show how the kinematic constraints can be obtained in the soft sector without violating the power counting. We further show how one can resum the threshold Sudakov logarithms systematically to all orders in a re-factorized framework with additional collinear-soft contributions. Direct applications to other small-$x$ processes involving heavy particles, jet (sub-)observables and EIC physics are straightforward.
\end{abstract}

\date{\today}

\maketitle

{\it Introduction}. 
In recent years, there have been growing interests in the small-$x$ physics at the RHIC and the LHC. For instance,  the suppression of
single inclusive hadron production in $dAu$ collisions at RHIC~\cite{Arsene:2004ux,Adams:2006uz} has
been suggested as the evidence for the gluon saturation at small-$x$ in the nucleus~\cite{JalilianMarian:2005jf,Albacete:2010bs}. In the small-$x$ region, where the longitudinal momentum fraction carried by the gluon becomes small, the gluon density grows dramatically and reaches the nonlinear regime. The color-dipole or the color glass-condensate (CGC) formalism~\cite{Gelis:2010nm,Mueller:1989st,Mueller:1993rr} are thus invoked and the linear BFKL equation~\cite{Balitsky:1978ic} is replaced by 
the nonlinear BK-JIMWLK equation~\cite{Balitsky:1995ub, Kovchegov:1999yj, JalilianMarian:1997jx, JalilianMarian:1997gr, Iancu:2000hn, Ferreiro:2001qy}, which inevitably leads to the gluon saturation~\cite{Gribov:1984tu,Mueller:1985wy} with a characteristic scale $Q_s^2$ at small-$x$. The saturation scale $Q_s$ features the typical transverse momentum of the gluons and grows as $x \to 0$. 

When $Q_s \gg \Lambda_{\rm QCD}$,  the usual collinear factorization in perturbative QCD can be applied, where one treats the nucleon/nucleus as a dilute system of partons. Given that $\alpha_s(Q_s)$ is typically not small, calculations beyond the leading order (LO) are required. Various attempts have been made in realizing the next-to-leading-order (NLO) predictions for the small-$x$ physics~\cite{Dumitru:2005gt,Altinoluk:2011qy,Chirilli:2011km,Beuf:2011xd,Roy:2018jxq,Liu:2019iml}. The calculations achieved in the seminal work~\cite{Chirilli:2011km} confirms the CGC factorization for the inclusive hadron production at NLO by explicitly showing that the rapidity divergence has a form of the BK kernel, which they absorbed into the nucleus structure function. Further works~\cite{Kang:2014lha,Altinoluk:2014eka,Watanabe:2015tja,Ducloue:2016shw,Iancu:2016vyg} to improve the results are pursued. Beyond fixed order (FO), a threshold resummation which is believed to maintain the positivity of this cross section~\cite{Xiao:2018zxf} is out of reach with all known approaches. Other Sudakov logarithms are also identified through NLO calculations of massive particle production in high energy $pA$ collisions and resummed~\cite{Mueller:2012uf,Mueller:2013wwa}.  However, a unified framework in the perturbative region which systematically incorporates all these ingredients, FO, BK evolution, Sudakov resummation and the relation to the collinear factorization~\cite{Kang:2012vm,Stasto:2014sea}, is by-far absent. Such a framework is crucial for the ultimate precision test of the small-$x$ formalism at the future electron-ion collider (EIC)~\cite{Accardi:2012qut}.

In this work, we propose the observable originated power counting as an approach to such a framework. General power counting has shown its tremendous power in FO calculations~\cite{Beneke:1997zp}, observable dependent factorization and resummations~\cite{Bauer:2000yr,Bauer:2002uv}, and in the small-$x$ evolution (see e.g.~\cite{Gelis:2010nm}).
Here we apply the observable originated power counting to the small-$x$ processes. In other words, we apply power counting directly to the small-$x$ observables. We use single hadron inclusive production at forward region to demonstrate how this can help to re-organize the NLO calculation with self-consistent logics and provide new insights, and how Sudakov threshold resummation and BK evolution can be realized simultaneously within one single unified framework.

{\it Inclusive hadron production.} We take single inclusive hadron production in proton-nucleus collisions, $p(p_P)A \to h(y_h, p_{h\perp})X$, as the example where we measure the hadron rapidity $y_h$ and transverse momentum $p_{h\perp}$. 
 For illustrative purpose, we only focus on the partonic channel  $q(p) A \to q'(z, p_\perp') X$. The quark $q'$ fragments into the identified hadron, with $ p_\perp' = p_{h\perp}/\xi$ and $\xi$ the momentum fraction of the parton carried by the hadron, $z = \nbar\cdot p'/\nbar \cdot p$ and $p$ the momentum of the incoming quark. 
We assume that the proton is moving along the direction $n^\mu = (1,0,0,1)$ while the nucleus along $\nbar^\mu = (1,0,0,-1)$. 
We focus on the region where $\lambda = p'_{\perp}/\nbar\cdot p 
\sim \sqrt{ - t/s} \ll 1 $ which defines the forward scattering. 
Other channels will be presented in~\cite{futurework}.

We start with the momentum fraction $1-z \sim {\cal O}(1)$. The observable $p_{h,\perp} \ll \sqrt{s}$ yields $3$ different modes to contribute to the leading region of the cross section when $p'_{\perp}
\sim p'_{\perp} (1-z) 
 \ll \nbar\cdot p (1-z)  \sim  \nbar\cdot p$:
 \begin{enumerate*}  
\item  collinear mode whose momentum scales as $(\nbar \cdot p , n\cdot p, p_\perp ) \sim  \nbar\cdot p (1, \lambda^2, \lambda)$ and 
\item soft mode with 
$k_s \sim   \nbar\cdot p (\lambda,\lambda,\lambda)$. 
\item Glauber potential with 
$k_G \sim  \nbar\cdot p (0,\lambda^\alpha,\lambda)$ with positive $\alpha \le 2$~\cite{Rothstein:2016bsq}, carried by the gluon from the nucleus to provide the transverse kick through the potential 
$1/k_G^2 \sim 1/k_{G,\perp}^2$. 
 \end{enumerate*}
 We note that the virtuality of all the modes 
 are ${\cal O}(p'_\perp)$ and they all contribute at leading power.
 Consequentially, 
 when $p_{h\perp} \gg \Lambda_{\rm QCD}$, 
 the NLO cross section reads
\bea\label{eq:forward-fac}
&\frac{ \mathrm{d}\sigma  }{\mathrm{d}y_h \mathrm{d}^2p_{h\perp} }
=   \frac{1}{4\pi^2} \int \frac{ \mathrm{d}  \xi }{\xi^2} \,
\frac{\mathrm{d} x}{x}\,
z x f_{q/P}(x,\mu) D_{h/q}(\xi,\mu) \nn 
& \hspace{-2.ex} \times  \int \mathrm{d}^2 b_\perp \mathrm{d}^2 b_\perp' 
\mathrm{d}^2 x_\perp  \,
e^{ip_\perp' \cdot r_\perp }
\left\langle
 S_w^{(2)}(b_\perp,x_\perp)S_w^{(2)}(x_\perp,b'_\perp) \right\rangle_\nu  
 \nn 
&  \hspace{-2.ex}  \times 
\left( 
{\cal I}_\delta^{(3)}(z, b_\perp,b_\perp' ,\mu)   {\cal S}_\delta^{(3)}(b_\perp,b_\perp')
 \delta^{(2)}\left(x_\perp - b_\perp \right) \right. \nn 
  &  \hspace{-2.ex}   \left.   +
{\cal I}^{(3)}(z, b_\perp,b_\perp', x_\perp,\nu)  + {\cal S}^{(3)}(b_\perp,b_\perp', x_\perp,\nu) 
\right) 
+{\cal O}(\alpha_s^2)   
\,.   \hspace{-2.ex}
\eea
We expect that the NLO factorized form can be promoted to all orders, with $n$-point functions and $n>3$ arising at higher orders. The collinear coefficients ${\cal I}^{(n)}$'s together with the parton distribution function (PDF) $f_{q/P}$ and the fragmentation function (FF) $D_{h/q}$ form the $n$-point forward TMD functions. These collinear coefficients can be thought of the matching coefficients of the forward TMD functions to the collinear PDF and FFs when $p_{h\perp} \gg \Lambda_{\rm QCD}$.
Here, ${\cal S}_\delta^{(3)}$ and ${\cal S}^{(3)}$ are soft functions, which we have identified in the small-$x$ formalism and will be given below. Note that these soft functions are not to be confused with the $S_w^{(n)}$ in the nucleus. 

Here $z = \tau/x \xi$ with $\tau = \nbar\cdot p_h/\nbar\cdot p_P  = p_{h\perp} e^{y_h}/\sqrt{s}$, and $r_\perp = b_\perp' - b_\perp$. $S_w^{(2)}(b_\perp, b_\perp') = \frac{1}{N_c} {\rm Tr}\left[ W(b_\perp) W^\dagger(b_\perp') \right]$
 with $W = {\cal P} \exp\left[ ig_s \int \mathrm{d}[\nbar\cdot \! \! x] n\cdot \!\! A (\nbar\cdot \!\! x,b_\perp)  \right]$ and $\langle \dots \rangle_\nu$ indicates the CGC 
average over the nuclear wave function and its small-$x$ evolution to the rapidity scale $\nu$. We will show that in our framework this scale is related to the initial gluon rapidity in the nucleus target $Y\sim \ln\frac{1}{x_A}$ in a natural way. Here, the $b_\perp$, $x_\perp$ scale like ${p_\perp'}^{-1}  \sim Q_s^{-1}$. 
We have assumed the light cone gauge $\nbar \cdot A = 0$ which we will stick to throughout the work.
 
The NLO corrections to the collinear coefficients have been obtained in various works~\cite{Chirilli:2011km, Chirilli:2012jd, Liu:2019iml}, using the light-cone perturbation theory. Here we re-derive the full results using covariant Feynman rules with the effective collinear approximation~\cite{futurework}. Dimensional regularization is used for regulating the usual infrared and collinear divergences and an $\eta$-regulator~\cite{Chiu:2012ir} is used to deal with the remaining rapidity divergence, whose effect for the collinear sector is equivalent to replace the overall $\frac{1}{1-z}$ factor with $\frac{1}{(1-z)^{1+\eta} } \left( \frac{\nu}{\nbar \cdot p} \right)^\eta$, in which the rapidity scale $\nu$ is introduced and one will expands around $\eta =0$ before doing the $\epsilon$-expansion in the dimensional regularization. Systematic power expansion in $\lambda$ is required in loop and phase space integration. At the end of the day, we have the renormalization scale $\mu$ associated with the pole in $1/\epsilon$ and the rapidity scale $\nu$ with the $1/\eta$ pole. The  pole part of the collinear contributions to the cross section in Eq.~(\ref{eq:forward-fac}) is found to be
\bea\label{eq:coll-div}
& -   \frac{ \alpha_s N_c }{2\pi\eta}  \,  
  \left[ \frac{\nu}{\nbar \cdot p }\right]^\eta \frac{1}{\pi} 
  \left[ \frac{r_\perp^2}{{r_\perp'}^2 {r_\perp''}^2}  \right]_+ 
 \left\langle 
    S_w^{(2)}  S_w^{(2)}
  \right\rangle_\nu
  \delta(1-z) \,  \nn
 & -   \frac{\alpha_sC_F }{2\pi \epsilon }  {\cal P}_{qq}(z) \,\!
  \left[ 1+ \frac{1}{z^2}e^{i \frac{1-z}{z}  p_\perp'  \cdot r_\perp }  \right]
 \, \! 
   \langle S_w^{(2)}(b_\perp,b_\perp') \rangle_\nu 
  \,,     \!\!
\eea
where we have included $\nu$-dependent part. Here 
plus-distributions over vector domains, $ \int \mathrm{d}^d x f(x) {\cal D}_+(x) = 
\lim_{\epsilon \to 0}\int \mathrm{d}^{d-2\epsilon} x [f(x)-f(0)] {\cal D}(x,\epsilon)$ with requiring $\lim_{|x|\to \infty} {\cal D}(x) \to 0$ 
are used~\cite{Chiu:2012ir, futurework} and
$r_\perp' = b_\perp - x_\perp$, $r_\perp'' = x_\perp - b_\perp'$ and ${\cal P}_{qq} = \left( \frac{1+z^2}{1-z} \right)_+$ is the quark splitting function. 
The finite collinear terms are found to be
\begin{widetext}
\bea
{\cal I}_{\delta, fin.}^{(3)}   =  &  
\delta(1-z) 
-   \frac{\alpha_s}{\pi}   C_F
 \left[
{\cal P}_{qq}(z) \ln \frac{r_\perp^2{\mu}^2}{c_0^2}
 -(1-z) 
 + \left( - \frac{3}{2} \ln \frac{r_\perp^2{p_\perp'}^2}{c_0^2} +\frac{1}{2} \right)  \delta(1-z)
 \right] 
 \frac{1}{2} \left( 1+ \frac{1}{z^2}e^{i \frac{1-z}{z}  p_\perp'  \cdot r_\perp }  \right) \nn 
&
- \frac{\alpha_s}{\pi}\left(
C_F - \frac{N_c}{2}  
\right)  
\left[  
 \frac{1+z^2}{(1-z)_+ }  \tilde{I}_{21 }(z) 
  -
 2 \left(
  (1+z^2) \left( \frac{\ln(1-z)}{1-z} \right)
\right)_+
     \right]  \,, 
\eea
\end{widetext}
where $c_0 = 2e^{-\gamma_E}$, and   
\bea
&
{\cal I}_{fin.}^{(3)}   = 
  \frac{\alpha_s}{\pi} \frac{N_c}{2}
 \frac{1}{\pi}
  \left[ 
  \frac{1+z^2}{(1-z)_+}  \frac{1}{z} \, 
e^{i\frac{1-z}{z} p_\perp' \cdot r_\perp' } \,
\frac{r_\perp' \cdot r_\perp''}{
{r_\perp'}^{2} {r_\perp''}^{2}
}\, \right. \nn 
& \hspace{-1.5ex} \left.   +
\int_0^1 \mathrm{d}\xi
 \frac{1+\xi^2}{(1-\xi)_+} \,  e^{ i \xi p_\perp'\cdot r_\perp' }\, 
 \left[
 \frac{e^{-i p_\perp'\cdot r_\perp' } }{{r_\perp'}^2 }  \right]_+  \,
\delta(1-z)  \right]  \,. 
 \eea
Here following Ref.~\cite{Chirilli:2012jd}, we have defined
 \bea
\tilde{I}_{21} =   &  \, \! 
\int\frac{ \mathrm{d}^2x_\perp  }{\pi} 
\left[
e^{-i(1-z)p_\perp' \!\! \cdot x_\perp} \left(
\frac{x_\perp \cdot (zx_\perp - r_\perp)}{x_\perp^2  (zx_\perp - r_\perp)^2}  
- \frac{1}{x_\perp^2} 
\right) \right. \nn 
 & \hspace{13.ex}  \left. + e^{-ip_\perp'\cdot x_\perp} \left( \frac{1}{x_\perp^2} \right)
\right] \,. 
\eea
 There are several points worth noticing here. First, there exist both rapidity and collinear poles, where the latter is absorbed into the proton PDF and the hadron FF while the former one will be shown to cancel against part of the rapidity poles from the soft contributions. Note that our interpretation of the rapidity pole from the collinear sector is different from the previous literatures~\cite{Chirilli:2011km,Chirilli:2012jd} where it was absorbed directly into the nucleus distribution. However this led to some difficulties in relating the rapidity scale $\nu$ to the initial gluon rapidity $Y$ in the nucleus target, since obviously from the $\nu$-dependence in Eq.~(\ref{eq:coll-div}), the $\nu$ from the collinear sector should be $\nbar \cdot p  = x \sqrt{s}$, associated with the proton size. We will show that it is the additional soft contribution that converts the $\nu$ to generically be associated with $Y$. Some earlier attempts along this direction were given in~\cite{Kang:2014lha, Liu:2019iml}.

Secondly, 
we note that there are several potentially different kinematic scales involved in this calculation, which are $\nbar \cdot p (1-z)$, $p_\perp'$ and $p_\perp'(1-z)$. 
Here the first scale is related to the rapidity divergence. When $1-z \sim {\cal O}(1)$, $\nbar \cdot p (1-z) \gg p_\perp' \sim p_\perp'(1-z)$ and the resulting logarithms can be eliminated by the scale choices $\mu \sim p_\perp'$ and $\nu \sim \nbar \cdot p $, which set the natural scales for the collinear matching
from the point of view of forward TMD functions. 
However when $p_{h\perp}$ close to the kinematic boundary, $\tau \to 1$, $1-z \sim {\cal O}(\lambda)$ and $\nbar \cdot p (1-z) \sim p_\perp' \gg p_\perp'(1-z)$, no scale choices can eliminate all large logarithms and therefore a threshold resummation in $\ln(1-z)$ is required, as will be discussed later. 

\begin{figure}
\hspace{6.ex}
\begin{subfigure}[c]{0.4\linewidth}
\includegraphics[width=\linewidth]{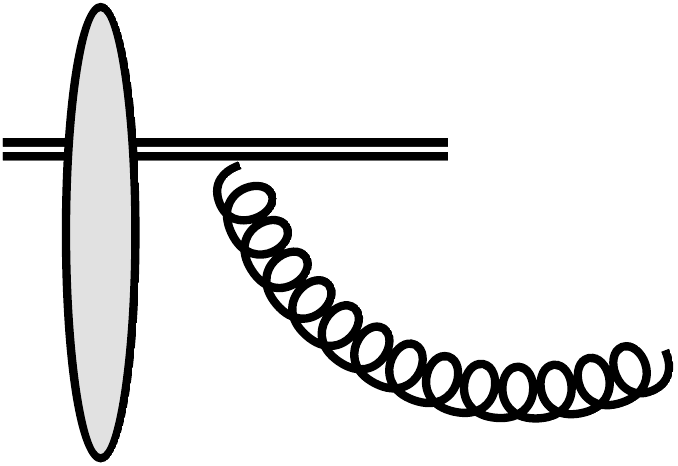} 
\end{subfigure}
\hfill
\begin{subfigure}[c]{0.4\linewidth}
\includegraphics[width=.83\linewidth]{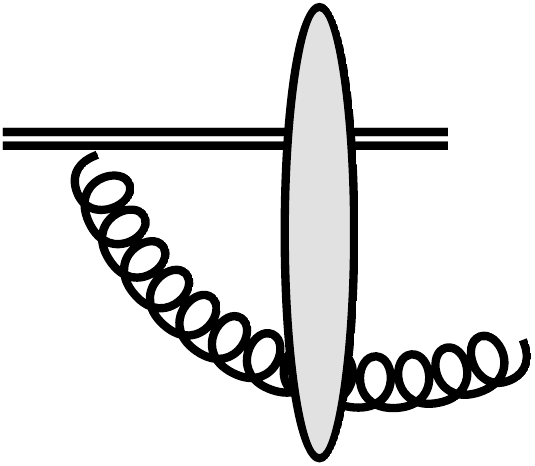} 
\end{subfigure}%
\caption{Eikonal vertex. Double line represents the eikonal line.}
\label{fig:eikonal-real}
\end{figure}

Now we turn to the soft sector. The NLO soft function can be calculated using the eikonal approximation dressed by the transverse momentum flow from the external source, as shown in Fig.~\ref{fig:eikonal-real}.
For the diagram on the left in Fig.~\ref{fig:eikonal-real}, we have for the small-$x$ soft current be
\bea
\hspace{-3.ex} 
{\cal J}_{\alpha,l}^a \! = \!  \frac{-g_s n_\alpha}{n\cdot k } 
\left(\frac{\nu}{k_\perp e^{|y|} }\right)^{\frac{\eta}{2}} 
 e^{i ( p_\perp'  + k_\perp) \cdot b_\perp}  
\left[ t^a W(b_\perp) \right]_{ij} \,, \hspace{-3.ex} 
\eea
where $k$ is the momentum carried by the soft gluon and $y$ is its rapidity. We have used the $\eta$-regulator to regulate the rapidity divergence in the soft eikonal current following~\cite{Chiu:2012ir} and~\cite{Liu:2013hba}. The diagram on the right gives 
\bea
{\cal J}^\alpha_{a,r}  &  =  -2 g_s
e^{-\frac{\eta}{2} |y| }
 \int \frac{\mathrm{d}^{d-2}l_\perp}{(2\pi)^{d-2}} 
\frac{l_\perp^\alpha}{l_\perp^2}  
\, \left( \frac{\nu}{l_\perp} \right)^{\eta/2} 
\\ 
& \hspace{ -4.ex} \times \int 
\mathrm{d}^{d-2} x_\perp 
e^{i p_\perp' \cdot b_\perp + i k_\perp \cdot x_\perp - i l_\perp \cdot r_\perp' }
\left[ W(b_\perp) t^b \right]_{ij} W_{ab}(x_\perp)\,.  \nonumber 
 \eea
Here $p_\perp'  \sim k_\perp \sim l_\perp \sim {\cal O}(\lambda)$  and $-l_\perp$ is the transverse momentum carried by the internal gluon line. We keep explicitly the dependence on the external nucleus source.  

When computing  the loop and the phase space integrals, we need to perform a systematic power expansion in $\lambda$ and keep only the leading contributions. For instance, 
for the momentum conservation, we will have $\delta(\nbar \cdot p - \nbar \cdot p' - \nbar \cdot k) \approx  \delta(\nbar \cdot p - \nbar \cdot p' )$, since 
$\nbar \cdot p - \nbar \cdot p' \sim {\cal O}(1) \gg  \nbar \cdot k \sim {\cal O}(\lambda)$. 
The power counting leads to the pole in the soft contribution
\bea
 \frac{ \alpha_s N_c }{\pi \eta} \left( \frac{\nu}{
{p_\perp'}
} \right)^\eta 
\frac{1}{\pi}
\left[ 
 \frac{r_\perp^2}{{r_\perp'}^2 {r_\perp''}^2} 
\right]_+
\left \langle 
S_w^{(2)} S_w^{(2)}  
\right\rangle_\nu\,,
\eea
and the finite term is found to be  
${\cal S}_{\delta,fin.}^{(2)} = 1$ and 
\bea
{\cal S}_{fin.}^{(3)}  = &  \frac{\alpha_s}{\pi}  \frac{N_c}{2}  \frac{1}{\pi}   \left(
\frac{ \ln({r_\perp'}^2 {p_\perp'}^2
/c_0^2)  
}{{r_\perp'}^2} 
+   \frac{ \ln({r_\perp''}^2 {p_\perp'}^2 
/c_0^2)  }{{r_\perp''}^2} 
\right. \nn 
& +  \left.  \frac{2 r_\perp' \cdot r_\perp''}{ {r_\perp'}^2 {r_\perp''}^2 }  \ln \left( \frac{ {r_\perp'}{r_\perp''}
{p_\perp'}^2
}{c_0^2}   
 \right)
\right)_+  \,.  
\eea
Here we note that  the finite terms arising in the soft function reproduce the kinematical constraint found in Ref.~\cite{Watanabe:2015tja}. 
We notice the similar BK structure in the soft rapidity pole as in the collinear sector.
 
The full NLO hadron production cross section obtained in a series of works~\cite{Chirilli:2011km, Chirilli:2012jd, Watanabe:2015tja, Liu:2019iml} can be produced by adding up the NLO collinear and soft results. We are left with the rapidity structure 
\bea\label{eq:raplog}
   &   \frac{\alpha_s N_c}{2\pi}   \delta(1-z) \left(
  \frac{1}{\eta}
- \ln\frac{p_\perp'^2}{\nu \nbar\cdot p }   
\right)   \times \nn  
&  
\int 
\frac{\mathrm{d}^2x_\perp}{\pi}
 \left[ \frac{r_\perp^2}{{r_\perp'}^2{r_\perp''}^2}  \right]_+
\!\!  \left\langle
 S_w^{(2)}(b_\perp,x_\perp)
 S_w^{(2)}(x_\perp,b_\perp') \right\rangle_\nu
  \,. \hspace{-2.ex}
\eea
The remaining rapidity pole will be absorbed into the
non-perturbative small-$x$ nucleus distribution and yields the ``correct'' BK equation by requiring the cross section is independent of the $\nu$: the resulting rapidity scale would be {\it correctly} associated with the rapidity scale for the initial gluons in the nucleus target. 
Such a rapidity scale is determined
by minimizing the logarithm in Eq.~(\ref{eq:raplog}), which generically sets $\nu = \frac{{p_\perp'}^2}{\nbar\cdot p}
 =  [n\cdot p_g]_{\rm LO}  = e^{-Y } [p_{g\perp}]_{\rm LO}/ \sqrt{s}  = x_A \sqrt{s}
$, with $p_g$ and $Y$ the momentum and rapidity of the gluon from the nucleus, respectively. 
Here we note that since $\ln\nu$ is always accompanied with the $\delta(1-z)$ term, only the LO kinematics need to be considered. $[n\cdot p_g]_{\rm LO}$ at LO is the minimum $+$-momentum can be reached by the initial gluon from the nucleus.
From this, we see how the rapidity scale is related to $Y$ and thus the evolution
$\mathrm{d}S_w^{(2)}/\mathrm{d} Y = - \mathrm{d}S_w^{(2)}/\mathrm{d}\ln\nu$.

To enclose this session, we note that when $p_{h\perp} \gg Q_s $, especially when $p_{h\perp} \sim  \nbar\cdot p$ a different power counting should be performed. Its connection to the collinear factorization can then be studied systematically and will be presented in a future work~\cite{futurework}. The different power counting also indicates the importance of the power suppressed terms and the results obtained should be matched with a collinear factorization calculation~\cite{Stasto:2014sea,futurework}.

{\it Threshold resummation}. As we noted before, when $1-z \sim {\cal O}(\lambda)$ there exist large threshold logarithms of $\ln(1-z)$, requiring resummation. 
In this case, where
$p'_{\perp}(1-z) \ll p'_{\perp} \sim  \nbar\cdot p(1-z) \ll  \nbar\cdot p $, energetic collinear real emissions are forbidden but soft radiations with momentum
$k_s \sim p'_{\perp}$ are still allowed. 
Now due to the threshold constrain $1-z \sim {\cal O}(\lambda)$, additional collinear-soft mode arises with the momentum
$
(\nbar \cdot  k_{cs}, n\cdot k_{cs}, k_{cs,\perp} )
\sim \nbar \cdot p \times \lambda \times (1,\lambda^2,\lambda)$ where the 
$\nbar \cdot k_{cs}$ component is sensitive to the threshold condition. 
To the leading power in $\lambda$, 
 the re-factorization of the cross section reads
 \bea\label{eq:thresholdfact}
 \mathrm{d}\sigma_{thresh.}
 = \sigma_{virt. + soft} \otimes_z S_{cs}(1-z) \,.
 \eea
 Here power corrections vanishing as $z \to 1$ are neglected and can be recovered by matching 
 Eq.~(\ref{eq:thresholdfact}) with the calculation in the previous section. 
The idea behind the factorization is that in the threshold limit, the physics at the short distance of order $1/p_\perp'$ is disentangled from the one with longer wave length of order $1/(p_\perp' \lambda)$. 
 The cross section is thus matched onto a collinear-soft theory, as shown in Fig.~\ref{fig:factorization-2}, with the ``hard" Wilson coefficient $\sigma_{virt. + soft} $ containing the collinear virtual, soft virtual and real corrections. The threshold PDF, FF as well as the nucleus distribution are included in the $\sigma_{virt. + soft} $. 
\bef
\includegraphics[width=.88\linewidth]{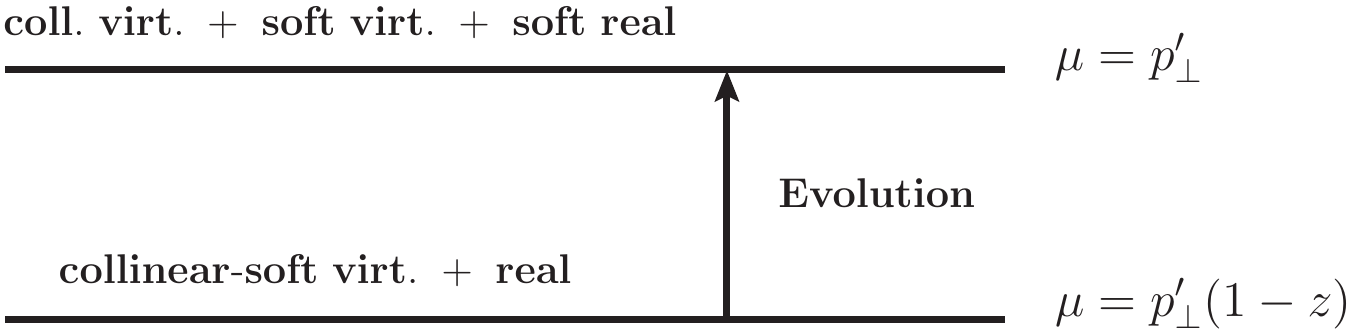} 
\vspace{2.ex}
\caption{Factorization scheme for $(1-z) \sim {\cal O}(\lambda)$. 
The threshold cross section is factorized into  the ``hard" Wilson coefficient $\sigma_{virt.+soft}$ with all virtuality of order $p_\perp'$ and a collinear-soft function $S_{cs}$.}
\label{fig:factorization-2}
\eef

For evaluating the Wilson coefficient $\sigma_{virt. + soft}$, the collinear and soft virtual calculations are identical to 
the previous section. 
 For the soft real emissions, 
 the $\nbar \cdot k$ component is restricted through the forward region threshold condition
 $\delta(1-z - \frac{\nbar \cdot k}{\nbar \cdot p} ) \theta(\nbar \cdot k - n\cdot k) + 
 \theta( n\cdot k- \nbar \cdot k ) 
 $ to be inserted into the real phase space calculations~\cite{futurework}.
  
 The diagrams for the collinear-soft function are topologically identical to squaring the soft currents in Fig.~\ref{fig:eikonal-real} with cuts over all possible lines. 
The squared amplitudes are now obtained by keeping the leading contribution as $ \lambda \sim (1-z) \to 0 $ and we note that $p_\perp'(1-z) \ll p_\perp'$. 
 For instance, when deriving the collinear-soft approximation, we utilized $e^{i(p_\perp' + k_{cs,\perp}) \cdot b_\perp } \approx e^{i p_\perp' \cdot b_\perp }$ and for the $b_\perp$, $x_\perp$ occur in $S_{\nu}^{(2)}$, they scale like ${p_\perp'}^{-1} $ and the conjugate momentum scale as $p_\perp'$. We thus find that the NLO collinear-soft function is fully decoupled from the external nucleus source and only contributes to the $\delta(x_\perp-b_\perp)$ term. 
 Also we note that when we encounter the propagator $(p' - k_{cs})^2$, the power counting goes like $(p' - k_{cs})^2 
 = \frac{1}{2} \nbar \cdot p  n\cdot k_{cs} 
 + \frac{1}{2} \nbar \cdot k_{cs}  n\cdot p
 + p_\perp \cdot k_{cs,\perp}  
 \sim \lambda^3 + \lambda^3 + \lambda^3$, therefore all three components contribute at the same power, which is different from the soft case in which 
 $(p' - k_s)^2 
 = \frac{1}{2} \nbar \cdot p  n\cdot k_s 
 + \frac{1}{2} \nbar \cdot k_s  n\cdot p
 + p_\perp \cdot k_{s,\perp}  
 \sim \lambda  + \lambda^3 + \lambda^2  \sim \frac{1}{2} \nbar \cdot p  n\cdot k_s 
 $. 
We found that the collinear-soft virtual corrections are scaleless and the real bare contribution gives 
 \bea
 S_{cs} =  - 
    \frac{ 2 \alpha_s  }{ \pi }  
  \frac{  \left( C_F  -   \frac{N_c}{2} \right)  }{(1-z)^{1+2\epsilon}}  
  \left(
  \frac{1}{\epsilon}-\frac{\pi^2}{12} \epsilon
  \right)
 \, 
  \left( \frac{\mu^2}{{p_\perp'}^2} \right)^{\epsilon}  \,,  
 \eea
 from which it is found that the natural scale for the collinear-soft function is 
 $\mu \sim p_\perp'(1-z)$ as expected. 
 It has been checked that all the poles cancel between $ \sigma_{virt. + soft}$ and the collinear-soft function
 and we fully reproduce the $1-z$ singular terms in the previous calculations, as a validation of the factorization formula in Eq.~(\ref{eq:thresholdfact}). The ${\cal O}(k_\perp/(1-z)\nbar \cdot p)$ terms, which were power suppressed when $1-z \sim {\cal O}(1)$, now arise. Only within our refactorization, such terms can be naturally captured. 
 From the $\epsilon$-poles of the bare results, 
 we extract its anomalous dimension 
 \bea
 \gamma_{cs} [\alpha_s] =4 \Gamma[\alpha_s] \left(
 C_F - \frac{N_c}{2}
 \right) \ln \frac{\mu {\bar N}}{p_\perp'} + \gamma[\alpha_s] \,,
 \eea
 where we have made the Mellin transformation 
 $f(N) = \int_0^1 \mathrm{d}z z^{N-1} f(z)$ and introduced the notation
 ${\bar N} = N e^{\gamma_E}$. Here, we have
 \bea
& 
 \Gamma^{(0)} = \frac{\alpha_s}{\pi} \,,  \qquad\qquad \gamma^{(0)} = 0
 \,, \nn
 & 
 \Gamma^{(1)} = \left( \frac{\alpha_s}{\pi} \right)^2
 \frac{1}{4}\left[ \left( \frac{67}{9} - \frac{\pi^2}{3} \right)C_A - \frac{20}{9}T_F n_F
 \right]\,. 
 \eea
The threshold logarithms can therefore be resummed by solving the renormalization group equation of the collinear-soft function 
$\mu \frac{\mathrm{d} S_{cs}}{\mathrm{d} \mu } = \gamma_{cs} S_{cs} \,,$ with the initial condition be the $\overline{\rm MS}$-renormalized collinear-soft function evaluated at $\mu = p_\perp'/{\bar N}$. The next-to-leading logarithmic collinear-soft function evolved to $\mu = p_\perp'$ reads
\bea
S_{cs} & \left( p_\perp' \right)
 =  S_{cs}\left(p_\perp'/{\bar N} \right) \nn 
&\times  e^{ - 
\int_0^1 \mathrm{d}z \, \frac{z^N-1}{1-z}
\int^{{p_\perp'}^2}_{{p_\perp'}^2(1-z)^2 } 
\frac{\mathrm{d}k^2}{k^2}  \Gamma[\alpha_s(k^2)] \left(
2 C_F -  N_c
\right)
} \,. 
\eea
The detailed derivations will be presented in~\cite{futurework}.

{\it Conclusions and outlooks}. In this work, we stress the importance of the observable originated power counting in the small-$x$ physics, which introduces novel soft contributions normally absent in the previous literatures and found to play important roles. We used the inclusive hadron production in $pA$ collision as an example to illustrate that within the soft function, one can systematically derive the kinematic constraints and still maintaining the factorization power counting. The rapidity scale $\ln \frac{1}{x_A}$ for the nucleus distribution naturally arises in our approach by requiring to minimize the logarithms when one adds up the collinear and the soft contributions. We also show that when near the kinematic boundaries, how the threshold Sudakov logarithms can be automatically resummed in the small-$x$ case by a re-factorization between the soft and the collinear-soft degrees of freedom. We show explicit results for resumming the threshold logarithms to the NLL accuracy. Last we note that our calculation is set up within the covariant Feynman rules and implements the complete dimensional regularization. This approach has the advantage being able to extend to higher orders with the aid of modern amplitude techniques~\cite{Li:2016ctv}. 

Our power counting scheme is expected to open up opportunities to the precision predictions of many other interesting small-$x$ processes. For instance, consistent treatments of the Sudakov resummation when heavy particles are present in the small-$x$ saturation formalism~\cite{Mueller:2013wwa} will become straightforward in the power counting framework and the resummation accuracy can be improved systematically. Other applications include studying the TMD physics, jet and its sub-observables in the small-$x$ regime, which we believe that taking into account suitable soft degrees of freedom incorporated in our framework are mandatory for achieving reliable theoretical predictions. Last we note that our approach is naturally applied to the EIC related studies to allow for relevant precision predictions.

{\it Acknowledgements.}  We thank H.~T.~Li, Y.~Q.~Ma and F.~Yuan for useful discussions. This work is supported by the National Science Foundation under Contract No.~PHY-1720486 (Z.K.) and by the National Natural Science Foundation of China under Grant No.~11775023 and the Fundamental Research Funds for the Central Universities (X.L.). 


\end{document}